\documentstyle[epsfig]{mn}
\begin{document}

\title
[Heating of the intergalactic medium due to structure formation
]
{Heating of the intergalactic medium due to structure formation}

\author[Biman B. Nath and Joseph Silk]
{Biman B. Nath$^{1,2}$ and Joseph Silk$^1$\\
$^1$Nuclear and Astrophysical Laboratory, University of Oxford, Keble Road,
Oxford OX1 3RH, UK\\
$^2$Raman Research Institute, Bangalore 560080, India
}
\maketitle

\begin{abstract}
{We estimate the heating of the intergalactic medium due to shocks
arising from structure formation. Heating of the gas outside the collapsed
regions, with small overdensities (${n_b \over {\bar n_b}}\ll 200$)
 is considered here, with
the aid of Zel'dovich approximation. We estimate the equation of state
of this gas, relating the density with its temperature, and its evolution
in time, considering the shock heating due to one-$\sigma$ density peaks
as being the most dominant. We also estimate the mass fraction
of gas above a given temperature as a function of redshift. We find that
the baryon fraction above $10^6$ K at $z=0$ is $\sim 10 \%$. 
We estimate the integrated Sunyaev-Zel'dovich
distortion from this gas at present epoch to be of order $10^{-6}$.
}
\end{abstract}


\section{Introduction}

It has become evident from recent numerical simulations  that a significant
fraction of the baryons in the universe reside in the warm-hot phase of the
intergalactic medium (WHIM), with temperatures of order
$10^5\hbox{--}10^7$ K (Cen \& Ostriker 1999, hereafter CO99).
The gas in this phase is raised to high temperature
by shock heating due to formation of structure. Recent simulations by
Dav\'e et al.~(2000)  and Croft et al.~(2000) have also calculated that
the equation of state of this phase
is approximately $\rho \propto T$.

There have been a few analytical attempts too to understand
the heating process in the intergalactic medium through analytic means. Pen
(1999) pointed out that there is a need for non-gravitational heating in the
intra-cluster and intergalactic medium (hereafter IGM) to avoid the constraints
from the soft-X-ray background. Extra heating decreases the amount of
clustering of the gas and therefore reduces the flux of soft X-ray radiation. 
Wu, Fabian
\& Nulsen (1999) have also addressed the question with detail
calculations and concluded the same.

These are, however, relevant for heating of the gas which are already within
collapsed objects. For example, the work of Wu et al (1999) refers to the
heating of the gas which is already within a collapsed halo, with overdensities
larger than $\sim 200$. The numerical simulations of CO99 and Dav\'e et al.~
(1999), on the other hand,
points out the heating in the gas which have overdensities much smaller
than this.

It is interesting to note that
Zel'dovich and his colleagues had reached similar conclusions to
that of the recent numerical simulations in the context of 
their study of formation
of pancakes. As a byproduct of their study of the formation of large pancakes,
they had worked out the magnitude of gravitational heating of the intergalactic
gas. Although much of the earlier motivation has been lost now, a substantial
part of their work sounds prescient. To quote from Sunyaev and Zeldovich
(1972)---``{\it 
It is possible that a significant fraction of the intergalactic gas
(10-50 \%) was not subjected to compression in the `pancakes' and was heated
only by the damped shock waves moving away from them.}'' This is exactly what
the numerical simulations have unearthed, namely, the heating of the gas
which with overdensities smaller than $\sim 200$, outside the collapsed
region but worked upon by shockwaves due to gravitational collapse.

In this {\it Letter},
 we attempt to understand the heating of this phase of IGM with the help
of Zel'dovich approximation. Since the gas in warm-hot IGM is only mildly
non-linear, this approximation can shed light on the gravitational heating
process, if used within its limitations. Below, we attempt to estimate the
amount of the gravitational heating, and the sate of the gas, by including
other heating and cooling effects. We also attempt to estimate the mass
fraction of baryons which are affected by this heating as a function of
redshift.

We assume a cosmological model with $\Omega_{\Lambda}=0.7$, $\Omega_m=0.3$
and $h=0.65$, with $\Omega_B \, h^2=0.015$, the big bang nucleosynthesis
value.

\section{Shock heating in the vicinity of collapsed objects}

Consider the gas surrounding a high density peak. As the gas flows inwards,
it is compressed, and depending on its adiabatic exponent, it stops at a place
away from the centre of accretion, and a shock wave travels outward. We
will concentrate on this shockwave as it compresses and heats the very
outer parts of the collapsed region. 
Sunyaev \& Zel'dovich (1972; hereafter SZ72) tried to model this shock wave in the context
of one-dimensional collapse of gas onto pancakes. In their idealized picture,
 as the gas
flows towards the inner region, a singularity appears and a shock
travels outwards through the gas. This shock velocity can be easily
determined for the perturbation of a given lengthscale $\lambda (=
2 \pi/k)$, assuming a single sinusoidal perturbation. Suppose the singularity
appears at a redshift $z_c$. They defined
a parameter, $\mu$ which corresponds to a given Lagrangian coordinate,
and is given by ${\sin \pi \mu \over \pi \mu}={1+z \over 1+z_c}$. The
parameter $\mu$, therefore, is equivalent to a time parameter.
In the case of a sinusoidal perturbation, it also gives the fraction of
matter that has passed through the shock wave up to a given moment.

\begin{figure}
\begin{center}
{\vskip-4mm}
\psfig{file=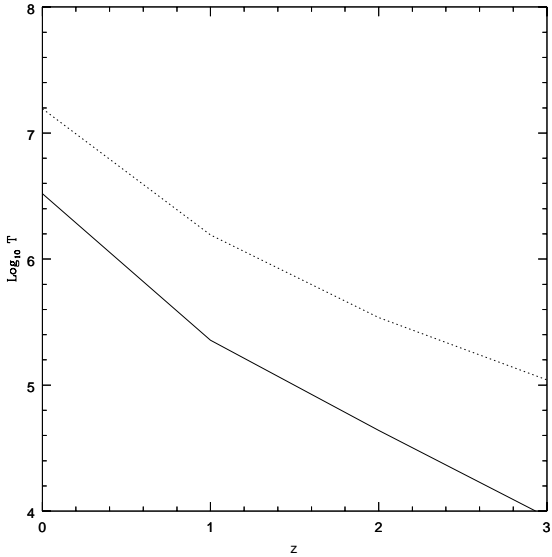, width=8cm}
\end{center}
{\vskip-3mm}
\caption{The maximum temperature from shocks due to structure formation
is shown as a function of redshift (solid line). The dotted line is that
from CO99.
}
\end{figure}

The velocity of matter falling onto the shock, $V_s$, as 
derived by SZ72, can be generalized for
any cosmological model as,
\begin{eqnarray}
V_s &\sim & {dz \over dt} {\lambda \over 2 \pi} {1 \over (1+z_c)^2}
(\mu \pi)^{1/2} \sin ^{1/2} (\mu \pi) \nonumber\\
&\sim &{dz \over dt} {\lambda \over 2 \pi} {1 \over (1+z_c)^2}
(\mu \pi) \,,
\end{eqnarray}
where $\lambda$ is the comoving lengthscale of perturbation.
The temperature behind the shock wave is given by $T_s \sim V_s^2 m_p/(6 k_B)$,
where $m_p$ is the mass of a proton and $k_B$ is the Boltzmann's constant
(SZ72, eqn(2)).

Note that this temperature reaches a maximum at $\mu \sim 0.5$, when
approximately half of the matter has passed through the shock. This
happens when $1+z \sim (2/\pi) (1+z_c)$. The maximum temperature is
given by, (noting that ${dz \over dt}=H(z) \,(1+z)$)
\begin{eqnarray}
T_{max}& \sim & {m_p \over 6 k_B} H(z)^2 \Bigl ({\lambda \over 2 \pi} \Bigr )^2
(\mu \pi)^2 \Bigl ({1+z \over 1+z_c} \bigr )^2 {1 \over (1+z_c)^2} \nonumber\\
&\sim & {m_p \over 6 k_B} H(z)^2 {L_{ln}^2 \over (1+z_c)^2} \,
\end{eqnarray}
where we have written $L_{ln}=1/k$ for the comoving lengthscale of the
perturbation, in the notation of CO99. This is the typical length of
perturbations that goes non-linear
at $1+z_c$. It is interesting to compare the Eqn (4) CO99 with this equation.
They derived a value of $K=0.3$ from their simulation where the maximum
temperature or, equivalently, the maximum sound velocity was given by
$C_s^2=K H^2 (L_{ln}/(1+z_c))^2$. 
Comparing with the above expression, we obtain,
$K \sim 5/18$ for a monoatomic gas.

There is, however, a crucial difference. The parameter $L_{ln}$ in CO99
is defined as the perturbation that goes non-linear at a given redshift z.
This provided the maximum temperature reached by the gas at a
given z. In the above formulation, however, there are three important
epochs: $z_{ln}$ is the epoch when the perturbation has an  overdensity larger
than unity and it goes non-linear, $z_c$ is the epoch
 when the singularity appears and
$z_m$ is the epoch
when $\mu \sim 0.5$, when the maximum gas temperature is achieved.
Naturally $z_{ln} > z_c > z_m$.  Here we have assumed that $z_{ln} \sim z_c$
but that it is larger than $z_m$.
This difference becomes non-negligible especially
 at high redshifts. This difference
is shown in Figure 1, where the maximum temperature reached
at a given redshift is plotted for the
$\Lambda$ cosmological model (as in CO99). The solid line is the
prediction from the above formulation (with $z_{ln}\sim z_{c} > z_m$
and the dotted line is from CO99.
The only difference is that we have treated $L_{ln}$ as the lengthscale
of the perturbation that goes non-linear at a redshift slightly larger
than $z_m$ to give the maximum temperature at $z_m$. Naturally, there is a
difference in these two epochs, of order of a Hubble time.

\section{Evolution of temperature}

In order to calculate the evolution of temperature of this gas, we will
need to take into account all sources of heating and cooling and their rates. 
Firstly, let us consider the rate of heating due to these shocks. We should
note here that the gas with different initial density (or overdensity) 
will go through the shockwave at different point of time.
Gas with larger initial density will be closer to the singularity and will
pass through the shock wave sooner. In the ideal case, the gas
density profile is smooth and shock travels through it equally affecting
all parts of it. In reality, however, we expect the gas closer to the
collapsed region will be shocked and heated more effectively than the gas
farther away.

We note that the average density profile of the gas away from the collapsed
region is expected to be of the type $\rho \propto r^{-2/3}$, where
$r$ is the perpendicular distance from the pancake or filament (Zel'dovich
1970). To treat the differential effectiveness of shock heating analytically,
 we shall define a timescale for the passing
of the shockwave through the gas as,
\begin{equation}
t_s \sim {\lambda \over 1+z_c} {1 \over U} \Bigl ({\rho \over
 {\bar \rho}} \bigr )^{-3/2} \,
\end{equation}
where $\bar{\rho}$ is the average ambient density, and where
$U=V_s/3$ is the velocity of the shock front relative to the
plane of symmetry (whereas $V_s$ is the velocity of matter impinging
on the shock (e.g., Jones, Palmer \& Wyse 1981)). The
extra factor $\Bigl ({\rho \over
 {\bar \rho}} \bigr ) ^{-3/2}$ then accounts for the fact that
gas with larger overdensity is heated more effectively than that with
lower overdensity. This prescription is valid only
for collapsing gas with different overdensities in a given perturbation.
In this work, we attempt
to calculate the equation of state of gas in a given 1-$\sigma$ perturbation.
We then estimate the equation of state of gas in the IGM in general using
the relevant filling factor of such perturbations.

We then write the shock heating rate as simply  $T_{shock}/
t_{shock}$. For a universe with $\Omega_{\Lambda}+\Omega_0=1$, one has,
\begin{eqnarray}
{dT_{shock} \over dz} &\sim &
-0.4 \times 10^6 \Bigl ( {\lambda /2 \pi \over 1 \, {\rm Mpc}} \Bigr )^2
(n_b/\bar{n_b})^{3/2} \nonumber\\
&& \times \mu ^3 {h^2 (1+z)^2 \over (1+z_c)^5}
 (\Omega_{\Lambda}+ (1+z)^3
\Omega_0) \, 
\end{eqnarray}

We should, however, remember the maximum temperature that the gas can be
raised to by the shocks, as discussed in \S 2. We therefore put an
upper limit on the temperature, as given by eqn(2). This will reflect
the physical fact that although lower density gas is not shocked
as effectively as the higher density gas closer to the filament or
pancake, higher density gas is not heated to indefinitely higher
temperatures this way.

As was pointed out in SZ72, a useful approximation for $\mu$ is
$\mu \sim {1 \over \pi} (6(1-{1+z \over 1+z_c}))^{1/2}$. We have
used this approximation in our calculations below.

Before discussing other sources of heating, we should note here that
this formulation is adequate only for a limited duration. Although in
principle, the gas infall continues until $\mu=1$, in reality, the
approximations used to calculate $\mu$ breaks down for large values
of $\mu$. We therefore consider the evolution of the temperature only
until $\mu =0.5$.

The second heating source, which is the adiabatic compression of the
gas, is easily described as, (for $n_b/\bar{n_b} \gg 1$)
\begin{equation}
{dT_{ad} \over dz}={2 \over 3} {T \over (n_b/\bar{n_b})} {d (n_b/\bar{n_b})
 \over dz}
\end{equation}
We characterize the growth of the overdensity by the following
equation,
\begin{equation}
{d (n_b /\bar{n_b}) \over dz}=-\eta {(n_b /\bar{n_b}) \over 1+z} \,,
\end{equation}
where $\eta$ equals unity for the linear regime in a $\Omega=1$
universe. In the quasi-linear regime, $\eta$ could be large.
For example, the overdensity evolves as $\delta \propto (1+z)^{-2.15}$ 
for the
range of the scales where the power spectrum has $n=-2$ (Peacock 1999).
We adopt a value of $\eta =2$. The final result is found not to depend
on its value strongly.

One cooling process
is due to the expansion of the universe, and is given by,
\begin{equation}
{dT_{ex} \over dz}= {2T \over 1+z} \,.
\end{equation}
Cooling due to free-free radiation is given by,
\begin{equation}
{dT_{ff} \over dz}=0.22 \Omega_b h {(n_b/\bar{n_b}) T^{1/2} (1+z_c)^2
\over (\Lambda _0 + (1+z)^3 \Omega _0)^{1/2}} \,.
\end{equation}
Although cooling due to inverse Compton scattering becomes important
at high redshift, at $z=4$, the cooling time for Compton cooling 
($\sim 9 \times 10^{12} (1+z)^{-4}$ yr $= 1.5 \times 10^{10}$ yr) is
larger than that for free-free cooling for a gas at $10^6$ K and with
an overdensity of $\sim 100$ ($\sim 10^9$ yr). It is shown below that
at high redshifts ($z_c \ga 3.5)$ shock heating contributes to the equation
of state only for gas with large overdensities, of order $\sim 100$.
Since Compton cooling is not as efficient as Compton cooling for this
gas, we neglect it in our calculation.

Combining all the heating and cooling processes, one has for the
evolution of the gas temperature,
\begin{equation}
{dT \over dz}={dT_s \over dz}+{dT_{ad} \over dz}+{dT_{ex} \over dz}
+{dT_{ff} \over dz}
\end{equation}
We present the numerical solution of this equation below.

\section{Results}

The process of heating due to structure formation is essentially
statistical in nature. To track it analytically, however, we focus
on the 1-$\sigma$ fluctuations. These are the perturbations that
dominate the heating at a given redshift, as found in CO99. Fluctuations
with higher degree of non-linearity at a given epoch would involve
gas with very large overdensities, and does not concern us here. As
has been emphasized earlier, here we are concerned with gas which is
rather on the outskirts of highly non-linear perturbations.

To calculate the equation of state of gas at a given epoch, we therefore
find out the relevant lengthscale, given the power spectrum which
is COBE normalized. For the physical state of the gas at a 
redshift $z$, we determine
the perturbation which goes non-linear at $z_c$, so that after evolving
by a time period equivalent to $\mu =0.5$, we reach the epoch $z$.
In other words, to find the state of the gas at
$z=0 (1, 2)$, we adopt $z_c=(1+z)*(\pi /2)-1 \sim .5 (2., 3.5)$.
We find the scale of the perturbations which are one-$\sigma$ at this
epoch, according to the relevant power spectrum. The value of
$L_{ln}$ at $z_c \sim 0.5 (2., 3.5)$ is $\sim 6 (1.6, 0.65) \, h^{-1}$ Mpc. 
The initial values of $\rho_b/{\bar \rho}$ is taken to be in the
range of $1\hbox{--}50$. The initial temperature has been fixed
at $3 \times 10^4$ K, which is the temperature reached by the IGM gas
through photoionization heating.
The results for the state of gas at $z=0,1,2$ are shown in 
Figure 2. The dotted line shows the equation of state as found by
Dav\'e et al~(2000).

\begin{figure}
\begin{center}
{\vskip-4mm}
\psfig{file=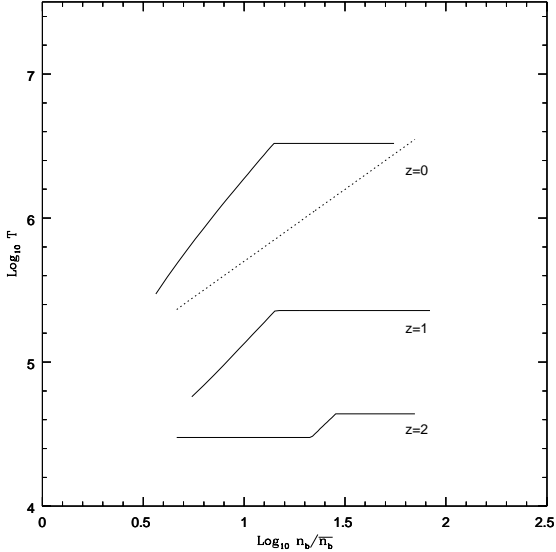, width=8cm}
\end{center}
{\vskip-3mm}
\caption{
Temperature of the gas is plotted against its overdensity $n_b/{\bar n_b}$.
The solid curves are for gas at $z=0,1,2$ from top to bottom. The dotted
line shows the equation of state found by Dav\'e et al.~(2000) (their 
Fig 6). 
}
\end{figure}

We can estimate the fraction of the mass that has a given temperature
in the following way. First, we note that, in the formulation of
SZ72, if the density of gas which has just entered the shock front is
$\rho_i$, then the fraction of mass ($f$)
 that has already gone through the shockwave
is given as, $(\rho_i / {\bar \rho}) \sim
3/(\pi ^2 f^2)$. If $T$ is the temperature of the gas corresponding to the
initial density $\rho_i$, then $f$ is the fraction of mass with temperature
larger than $T$. To be precise, this approximation is valid for the case
of instantaneous cooling, which is a reasonable assumption for $\mu \la 0.1$
($z_c \ga 0.5$, as can be seen from eqns (4) \& (9)). In other words, this
approximation is valid only for small values of $f (\ll 1)$.

We multiply this fraction with the (comoving) number density of the
one-$\sigma$ peak, as given in Bardeen et al.~(1986), to derive the
fraction of mass which has temperature larger than a given value.
The results for the fractions at $z=0,1,2$ are shown in
Figure 3. Unfortunately, the limitations of the single sinusoidal
wave approximation do not allow us to draw a full curve, as one has to stop
at $f \sim 1$ (in reality, the above approximation is valid only for small
values of $f$). The curves, however, can serve as pointers to what one-$\sigma$
peaks can do to the diffuse IGM.

We also show by stars the mass fraction derived by
Croft et al.~(2000) at $z=0, 2$, and the fractions are $41 \%$ and
$5 \%$ respectively. To compare these numbers with the curves in Figure 3, 
we should remember that the curves show the result of heating by one-$\sigma$
density peaks only. In reality, there will be contribution from higher sigma
peaks, taking the gas to higher temperatures at rarer places.
Although we do not have the fraction
for gas above $10^5$ K at $z=0$, a naive extrapolation of the existing
curve above $10^6$ K is consistent with this value. We note here that
the curve for $z=0$ shows that the fraction of gas above $10^6$ K is
of order $10 \%$. 

We note here that the mass fractions from the simulation of CO99 are
much larger than these. Since the simulations use different techniques and
employ different resolutions, it is not obvious to what do these discrepancies
owe their existence and if they are of much importance.

\begin{figure}
\begin{center}
{\vskip-4mm}
\psfig{file=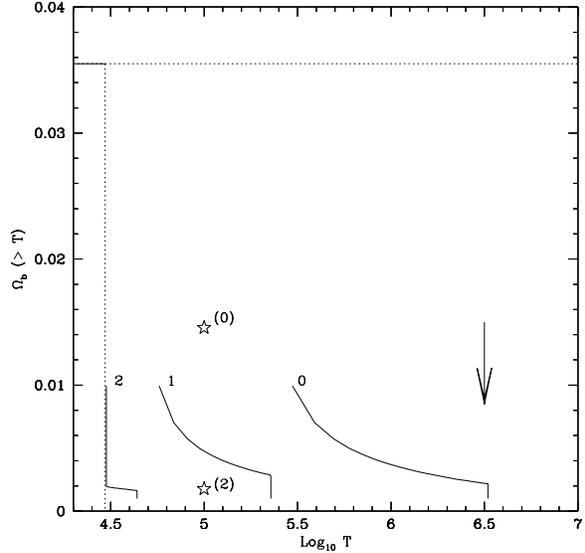, width=8cm}
\end{center}
{\vskip-3mm}
\caption{
The fraction of shock heated diffuse gas above a given temperature
$\Omega_B(>T)$ is plotted against logarithm of $T$. The three solid curves from
right to left are for gas at $z=0,1,2$ respectively. The stars
are the mass fraction at $z=0,$ and $z=2$ found by Croft et al.~(2000,
their \S 2).
The horizontal dotted line refers to the total baryonic gas in the universe
and the vertical dotted line shows the lower limit of the temperature
due to photoionization. The arrow is from the limit of soft X-ray
emission for gas with $n_b/\bar{n_b} \sim 100$ at $z=0$.
}
\end{figure}

\section{Discussions}

The equation of state for low density gas in Figure 2 depends on the
particular form of the timescale for passing of the shockwave in eqn (3).
as is evident from the analytical solution (eqn (10)). In reality, modelling
the efficiency of shock heating might add additional parameters and
the resulting equation would therefore have some scatter.

In Figure 3, we also plot the limits from the observations of the soft 
X-ray background.
After the subtraction of the discrete sources, it now appears that a
flux of $4$ keV cm$^{-2}$ s$^{-1}$ sr$^{-1}$ keV$^{-1}$ at $0.25$ keV can
be taken as an upper limit to any possible contribution from diffuse matter
in the IGM (e.g., Wu, Fabian \& Nulsen 1999). 
We apply this limit to our result for IGM at $z=0$. If the
gas at temperature $T$ (with the corresponding density $n(T)$, as given in
Fig 2 for $z=0$) has a filling factor $\epsilon_T$, then the flux at 0.25 keV
is proportional to $Flux \, (keV/cm^2 \, s \, sr \, keV)  
\propto  \epsilon_T n(T)^2 \, T^{-0.5}
\, \exp(- 0.25 \, keV/k_BT) \, (c/H_0) \,$
and this allows us to calculate $\epsilon_T$ as a function of $T$. Since
$\epsilon_T$ is found to decrease very quickly with $T$, we can approximate
$\epsilon (>T) \sim \epsilon_T$ and use it to put a limit on the mass fraction
of gas with temperature larger than a given value $T$. This limit is shown
as a dashed arrow in Figure 3 for gas with $n_b/\bar{n_b}=100$ at $T=T_{max}$
for $z=0$. The limit for lower density gas is less restrictive. 
The above limit is obviously
uncertain to the extent that the appropriate path length differs from
the approximate $c/H_0$.

In this calculation we have used a metallicity of $0.01$ solar abundance,
which is the metallicity found in the Lyman-$\alpha$ absorption systems
at high redshift, and is relevant for the diffuse gas considered here.
It is possible that the metallicity of this diffuse gas increases in time,
as found in the numerical simulation by Cen \& Ostriker (1999b). It is, 
however, not yet known from observations how this metallicity evolves in
time, and its dependence, if any, with the density and clumpiness of the gas.
Any metallicity will make the above limit more stringent; in other words,
make the contribution of the diffuse shock-heated IGM gas towards to the
soft X-ray background much more prominent. As the detail numerical
simulation of the
soft X-ray background radiation by Croft et al.~(2000) shows, the
contribution of the (enriched) shocked heated diffuse gas is comparable
to the upper limit of the unresolved X-ray emission in the soft band.

Given the relation between density and temperature, we can estimate
the resulting Sunyaev-Zel'dovich distortion on the cosmic microwave
background. Defining $f_{los}$ to be the fraction of the line of sight
going through hot filamentary structures, one can estimate the integrated
Compton $y$ parameter aas,
\begin{equation}
y \sim 10^{-6} \, \Bigl ({n_b/\bar{n_b} \over 10^2} \Bigr )
\,\Bigl ({T_e \over 10^6\, K} \Bigr ) \, \Bigl ({c/H_0 \over 10^3 \, Mpc}
\Bigr ) \Bigl ({f_{los} \over 0.03} \Bigr ) \,.
\end{equation} 
The (comoving) volume fraction of 1-$\sigma$ peaks from Bardeen et al.~(1986)
is of order $3 \%$.
Another way of estimating this is to use the fact that
the gas with density $\ga n_b$ occupies a lengthscale
${\lambda \over  (1+z_c)} (n_b / {\bar n_b})^{-3/2}$,
The Compton $y$-parameter is approximated as,
\begin{equation}
y \sim  {2 \lambda \over (1+z_c)} (n_b / {\bar n_b})^{-3/2} \, 
{n_b k_B T_e \over m_e c^2} \, \sigma_T \,,
\end{equation}
where $\sigma_T$ is the Thomson cross-section. For the case at $z=0$,
$y$ for a single structure
is found to be of order $\sim 4 \times 10^{-8}$ for $n_b/ {\bar n_b}
\sim 100$. There will be, however, of order
$((c/H_0)/L_{ln}) \times f_{los} \sim 30 (f_{los}/0.03)$  such
structures in one line of sight, and so the total distortion will
amount to $y \sim 10^{-6}$.

\section{Summary}

We have applied the Zel'dovich approximation to estimate
the heating of the diffuse intergalactic medium by
shocks associated with 1-$\sigma$ density peaks in structure formation
at different redshifts. We are able to reproduce the equation of state
of the warm-hot IGM found in recent numerical simulations.
We estimate the baryon fraction of the
gas above $10^6$ K at the present epoch to be at least $\sim 0.1 \Omega_b$.
The integrated Sunyaev-Zel'dovich distortion from the diffuse IGM
filaments amounts to $y \sim 10^{-6}$.

\bigskip
BN acknowledges joint support from the Indian National
Science Academy and the Royal Society, UK, and thanks the 
Astrophysics Department of the University of Oxford for hospitality.
We thank the anynomous referee for detail comments on the paper.

\end{document}